\documentclass[aps, prx, twocolumn, reprint, nofootinbib]{revtex4-1}

\usepackage{amsmath}    
\usepackage{mathrsfs}
\usepackage{graphicx}   
\usepackage{verbatim}   
\usepackage{subfigure}  
\usepackage{upgreek}
\usepackage{hyperref}
\begin{document}

\newcommand{\TEMOO}{TEM$_{00}~$}
\newcommand{\TEMIO}{TEM$_{10}~$}
\newcommand{\SiN}{Si$_3$N$_4~$}
\newcommand{\um}{$\upmu$m}
\newcommand{\opt}{_{\text{opt}}}
\newcommand{\eff}{_{\text{eff}}}
\newcommand{\tot}{_{\text{tot}}}
\newcommand{\mat}{_{\text{mat}}}
\newcommand{\unit}{_{\text{unit}}}
\newcommand{\others}[1]{}
\newcommand{\on}{_{\text{on}}}
\newcommand{\off}{_{\text{off}}}

\title{Optically Defined Mechanical Geometry}
\author{Abeer Z. Barasheed}
\author{Tina M\"uller}
\author{Jack C. Sankey}

\affiliation{McGill University Department of Physics}

\date{\today}

\begin{abstract}
	In the field of optomechanics, radiation forces have provided a particularly high level of control over the frequency and dissipation of mechanical elements. Here we propose a class of optomechanical systems in which light exerts a similarly profound influence over two other fundamental parameters: geometry and mass. By applying an optical trap to one lattice site of an extended phononic crystal, we show it is possible to create a tunable, localized mechanical mode. Owing to light's simultaneous and constructive coupling with the structure's continuum of modes, we estimate that a trap power at the level of a single intracavity photon should be capable of producing a significant effect within a realistic, chip-scale device.
\end{abstract}

\maketitle

Solid state mechanical systems are ubiquitous throughout society, from oscillators in time-keeping devices to accelerometers and electronic filters in automobiles and cell phones. They also comprise an indispensable set of tools for fundamental and applied science. For example, using tiny mechanical systems, it is possible to ``feel around'' surfaces at the atomic scale \cite{Hembacher2004Force}\others{Garcia2012The}, detect mass changes from adsorbed chemicals with single-proton resolution \cite{Chaste2012A}, and sense the gentle magnetic ``tugs'' from individual electron spins \cite{Rugar2004Single}, persistent currents in a normal-metal ring \cite{Castellanos2013Measurement}, or even element-specific nanoscale clusters of nuclei \cite{Degen2009Nanoscale}\others{Mamin2009Isotope, Poggio2010Force}. Meanwhile, human-scale masses (positioned kilometers apart) currently ``listen'' for gravitational waves emitted by violent events across the universe \cite{Abbott2016Observation}.

In the field of optomechanics, the forces generated by light provide a means of tuning the fundamental properties -- namely frequency, dissipation, normal mode geometry, and effective mass -- of mechanical systems at every size scale \cite{Aspelmeyer2014Cavity}. The frequency and dissipation have been particularly well-controlled, often tuned by many orders of magnitude using feedback techniques \cite{Cohadon1999Cooling}, bolometric effects \cite{Zalalutdinov2001Autoparametric, Vogel2003Optically, Metzger2004Cavity}, or radiation pressure \cite{Carmon2005Temporal, Kippenberg2005Analysis, Arcizet2006Radiation, Gigan2006Self, Schliesser2006Radiation, Ni2012Enhancement}. The geometry and mass have also been tuned via optically-mediated normal mode hybridization \cite{Lin2010Coherent, Safavi2011Electromagnetically, Zhang2012Synchronization, Massel2012Multimode, Ni2012Enhancement, Shkarin2014Optically, Fu2014Optically}, but not so profoundly: only a few (essentially two) normal modes are involved, and the resulting hybridized modes therefore exhibit a mass and spatial extent comparable to that of the unperturbed modes.

Here we propose to exploit radiation pressure's simultaneous influence over a \emph{continuum} of modes to strongly tune the geometry and mass of a mechanical system. The basic idea is to fabricate an extended phononic crystal structure \others{Kushwaha1993Acoustic}\cite{Pennec2010Two} and apply an optical trap to one lattice site, thereby creating a defect that exponentially localizes one or more mechanical modes. Unlike \emph{structurally} defined defect modes \cite{Pennec2010Two} -- realized some time ago \cite{Torres1999Sonic} and currently exploited with extraordinary success in optomechanics \cite{Alegre2011Quasi, Gavartin2011Optomechanical, Chan2011Laser, Safavi2012Observation, Hill2012Coherent, Safavi2013Squeezed, Yu2014A, Cohen2015Phonon, Meenehan2015Pulsed} -- we show that the spatial extent and mass of \emph{optically} defined defect modes can be tuned by many orders of magnitude using a realistic, chip-scale optomechanical geometry. Additionally, despite the comparatively weak optomechanical interaction with each of the unperturbed, extended mechanical modes, we estimate that an optical trap having an average intracavity power corresponding to a single photon should in principle cause a macroscopic, measurable change in the amplitude of a millimeter-scale mechanical element. Moreover, we show that a larger structure will exhibit a \emph{larger} response to a given trap, despite its larger mass. Section \ref{sec:1D} describes the basic physics with an analytical toy model in one-dimension (1D), section \ref{sec:2D} describes a an implementation based on existing fabrication and optomechanical techniques, and section \ref{sec:discussion} discusses some of the potential research directions enabled by this type of coupling. In particular, we suspect the ability to optically tune the spatial extent of a mechanical mode will provide a unique platform for fundamental dissipation studies, unconventional sensing applications, and quantum optomechanics experiments.


\section{Toy Model in 1D}\label{sec:1D}

\begin{figure}
	\includegraphics[width=1.0\columnwidth]{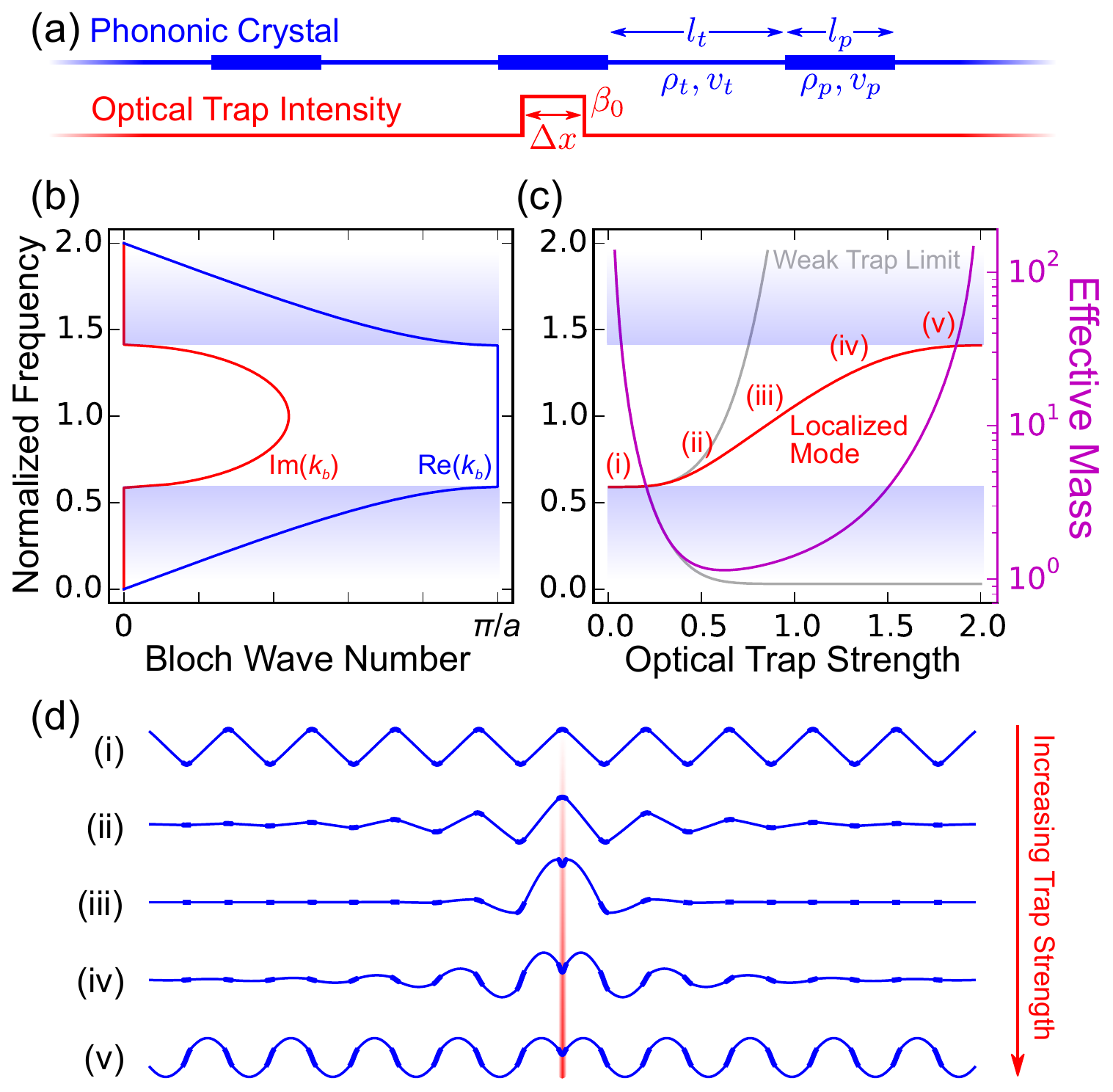}
	\caption{Optical localization in 1D. (a) Infinite phononic crystal string under tension $T$ with pad (tether) mass densities $\rho_p$ ($\rho_t$), lengths $l_p$ ($l_t$), and wave speeds $v_p$ ($v_t$). The unit cell mass $m\unit=\rho_p l_p+\rho_t l_t$ and length $a=l_p+l_t$. An optical trap of uniform spring constant density $\beta_0$ (red) is applied to the central region ($\Delta x$) of a single pad. (b) Dependence of real (black) and imaginary (red) components of the Bloch wave number $k_b$ on mechanical frequency $\omega$ for the unperturbed crystal. For this calculation, $l_t/l_p=4$ and $v_t/v_p=4$, and all frequencies are normalized by the mid-gap value $\omega_g = \pi/(l_t/v_t+l_p/v_p)$. (c) Dependence of localized mode frequency (red) and effective mass (magenta, normalized by $m\unit$) on integrated trap strength $\sqrt{\beta_0 \Delta x / m\unit}$ (i.e.~represented as the frequency of a rigid point mass $m\unit$ experiencing the same restoring force). Gray curves show the same calculation in the weak trap limit. (d) Optically tuned mode shapes for values labeled in (c). Red gradient qualitatively indicates the location and intensity of the trap.}
	\label{fig:basics}
\end{figure}

To gain some immediate intuition, we first consider an infinitely-long, ideal string under tension $T$, with periodically alternating mass density $\rho_p$ (for the heavier ``pad'' region) and $\rho_t$ (for the lighter ``tether'' region), as drawn in Fig.~\ref{fig:basics}(a). The optical trap is modeled as a transverse spring constant density $\beta(x)$ proportional to the average (classical) laser power or cavity occupancy $\bar{n}_\gamma$ \cite{Chang2012Ultrahigh, Muller2015Enhanced}, so that the string's wave equation becomes
\begin{equation}\label{eq:wave}
\rho\partial_t^2 y-T\partial_{x}^{2} y+\beta y = 0,
\end{equation}
where $\rho(x)$ is the local mass density at position $x$ and $y(x)$ is the transverse displacement. For simplicity, $\beta(x)$ is assumed to be a constant ($\beta_0$) within a region of length $\Delta x$ at the center of one pad and zero elsewhere, as illustrated in Fig.~\ref{fig:basics}(a). When $\beta_0=0$, the alternating mass densities $\rho_{p,t}$ (and resulting wave velocities $v_{p,t}=\sqrt{T/\rho_{p,t}}$) lead to Bloch-periodic solutions and the well-known dispersion relation between mechanical frequency $\omega$ and Bloch wave number $k_b$ shown in Fig.~\ref{fig:basics}(b) (with unit cell length $a=l_p+l_t$ and pad (tether) length $l_p$ ($l_t$)). For low and high frequencies (shaded bands, Im$[k_b]=0$), the structure supports a continuum of delocalized, propagating modes, while for frequencies in the gap ($k_b$ complex), propagation decays exponentially. Like an optical Bragg mirror, the ratio of the gap width $\Delta\omega_g$ to the mid-value $\omega_g$ increases with velocity contrast $v_t/v_p$, and is maximized when the wave transit times $t_{p,t}=l_{p,t}/v_{p,t}$ for the pad and tether are equal.

If $\beta_0>0$, the local wavenumber $k_t$ within the trapped region becomes
\begin{equation}\label{eq:trap-dispersion}
k_{t} =\frac{1}{v_{p}}\sqrt{\omega^{2}-\omega\opt^2}.
\end{equation}
where $\omega\opt^2 \equiv \beta_0/\rho_p$. In analogy with quantum mechanics, the trap here plays the role of ``potential barrier'' for phonons: low-intensity traps reduce $k_{t}$, and traps having $\beta > \rho_p \omega^2$ result in purely imaginary $k_{t}$ -- propagation is ``forbidden'', and only evanescent solutions exist within the trap. The boundary conditions at the edges of the trap yield a transcendental equation
\begin{equation}\label{eq:trans}
	\frac{\omega}{v_{p}}\frac{e^{i\omega t_{p}/2}-iA}{e^{i\omega t_{p}/2}+iA}=ik_{t}\tan(k_{t}\Delta x/2)
\end{equation}
with $A \equiv (v_+ e^{i\omega t_+}-v_- e^{i\omega t_-}+v_{tp} e^{ik_b a})/ 2\sin \omega t_{t}$, $v_\pm~\equiv~(v_{p}\pm v_{t})/(v_m \mp v_p)$, $v_{tp} \equiv 4v_{p}v_{t}/(v_{t}^{2}-v_{p}^{2})$, and $t_\pm \equiv t_p\pm t_t$. Solving Eq.~\ref{eq:trans} yields the eigenfrequency $\omega$ plotted (red) in Fig.~\ref{fig:basics}(c). Figure \ref{fig:basics}(d) shows the shape of this mode for the five trap strengths indicated in (c). As the trap strength is increased, the delocalized lower band edge mode (i) collapses (ii)-(iii) as its frequency enters the gap, and subsequently expands (iv) into another delocalized mode (v) that, outside the trapped region, looks identical to the untrapped upper band edge mode. 

In this way, a light field can dramatically alter the amount of material participating in a mechanical oscillation. To quantify this statement, we extract an effective mass $m\eff$ from the relationship between displacement $y(x_0)$ at a location $x_0$ and the total energy of the structure
\begin{equation}\label{eq:Utot}
U_\text{tot} = \frac{1}{2} \left[ \int \rho(x) \omega^2 \left(\frac{y(x)}{y(x_0)}\right)^2 dx \right ] y(x_0)^2.
\end{equation}
Since $y(x) \propto y(x_0)$ for all $x$, the integral as written is independent of $y(x_0)$, and we can identify an effective spring constant $K\eff$ as the quantity in square brackets. This, together with the known mechanical frequency $\omega$, defines $m\eff = K\eff/\omega^2 = \int \rho \left[y/y(x_0)\right]^2 dx$, which is plotted in Fig.~\ref{fig:basics}(c) for $x_0$ at the center of the trapped pad. This definition of $m\eff$ is intuitively consistent with the mode profiles in Fig.~\ref{fig:basics}(d), as well as the functional form of the localized mode frequency (c): the trap's ability to tune $\omega$ is largest when $m\eff$ is small (roughly\footnote{Note the deviation from this simple intuition arises from the choice of $x_0$ in Eq.~\ref{eq:Utot}. The trap essentially serves as a ``partially clamped'' region that suppresses the amplitude at $x_0$, leading to the ``dimples'' at high trap strength in Fig.~\ref{fig:basics}(d), a systematically larger $m\eff$, and a reduced trap efficiency.}). 

We emphasize that for this \emph{infinite} structure (i.e. having \emph{infinite} untrapped $m\eff$), even an infinitesimal trap will produce a finite value of $m\eff$. We can quickly illuminate this behavior by extracting analytical expressions for the trapped mechanical frequency $\omega$ and the localization length $\mathscr{L}\equiv 1/\text{Im}[k_b]$ for the case of an optimal crystal ($t_p=t_t$, with unperturbed band-edge frequency $\omega_0$) and a weak trap $\omega\opt \ll \omega_0$. Expanding Eq.~\ref{eq:trans} in the small parameters $\Delta \equiv \omega\opt/\omega_0$ and $\delta \equiv (\omega-\omega_0)/\omega_0$, the frequency shift
\begin{equation}
\delta \approx \frac{(v_{p}-v_{t})^{2}\left(\frac{\omega_{0}\Delta x}{v_{p}}+\sin\frac{\omega_{0}\Delta x}{v_{p}}\right)^{2}\tan\omega_{0}t_{p}}{32v_{p}v_{t}\omega_{0}t_{p}}\Delta^4
\end{equation}
and localization length
\begin{equation}
\mathscr{L} \approx \frac{4v_{p}v_{t}a}{\left(v_{t}^{2}-v_{p}^{2}\right)\left(\frac{\omega_{0}\Delta x}{v_{p}}+\sin\frac{\omega_{0}\Delta x}{v_{p}}\right)\sin(\omega_{0}t_{p})}\frac{1}{\Delta^{2}}.
\end{equation}
Finally, since $k_b \approx \pi/a + i/\mathscr{L}$ and $\mathscr{L}\gg a$ in this limit, the (piece-wise) integral for the effective mass reduces to a geometric series, and
\begin{equation}\label{eq:meff}
m\eff \approx m_{p}+\frac{2m_{p}+2m_{t}}{e^{2a/\mathscr{L}}-1} \approx(m_{p}+m_{t})\frac{\mathscr{L}}{a},
\end{equation}
where $m_{p} \equiv \rho_{p}l_{p}/2+\rho_{p}v_{p}\sin\omega_{0}t_{p}/2\omega_{0}$ and $m_t\equiv  (\rho_{t}l_{t}/2-\rho_{t}v_{t}\sin\omega_{0}t_{p}/2\omega_{0})\cot(\omega_{0}t_{p}/2)$ are the effective masses of the pad and tether for the unperturbed band edge mode. The frequency shift $\delta$ and the first (marginally more precise) expression of Eq.~\ref{eq:meff} are plotted (gray) in Fig.~\ref{fig:basics}(c). Importantly, $m\eff\propto\mathscr{L}\propto 1/\Delta^2$, meaning localization occurs for any strength of trap. Similarly, the transition back to infinite $m\eff$ in Fig.~\ref{fig:basics}(c) is not asymptotic, occurring at a finite value of $\beta_0$. 


\section{Realistic Implementation in 2D}\label{sec:2D}
\begin{figure*}[htb]
	\includegraphics[width=0.95\textwidth]{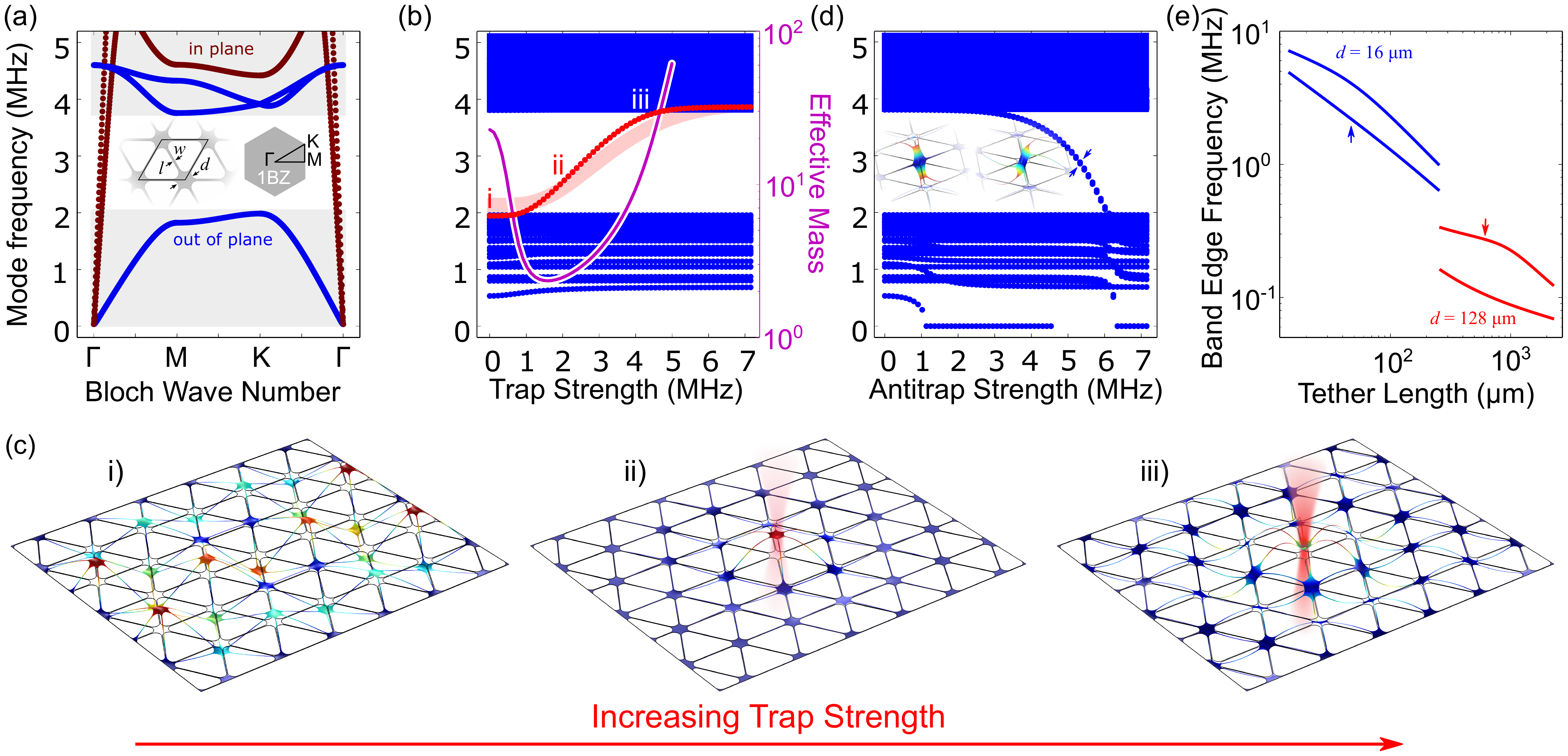}
	\caption{Finite-element (COMSOL) simulation of phononic pseudogap membrane, fabricated from 100-nm-thick \SiN having an internal stress of 1 GPa. (a) Dispersion for an infinite structure (unit cell inset) with dimensions $w = 1$ \um, $d = 16$ \um, and $l = 52.5$ \um. Red points show in-plane modes, and blue shows out-of-plane (OOP) modes. Right-hand inset shows first Brillouin zone (1BZ, gray) and $\vec{k}_b$-labeling convention. (b) Eigenfrequencies of a finite structure with an 8-\um-diameter optical trap applied to the central pad. Trap strength again scaled to the frequency of a single ridid unit cell trapped alone in outer space. Band edge mode (red, see (i) of (c) below) can be tuned through the OOP gap. Effective mass (magenta) calculated from COMSOL, plummets as the mode localizes. Faint red line shows the prediction of the 1D analytical model with linear mass densities and tension determined by the structure's periodicity in the K direction. (c) Mode profiles at the points indicated in (b), showing (i) untrapped mode, (ii) localization, (iii), delocalization. (d) Two nearly-degenerate modes (insets) localized by an optical anti-trap. (e) Dependence of band-edge frequencies on tether length for small $d=16$ \um (blue) and large $d=128$ \um~(red) structures with $w=1$ \um. At the optimal tether lengths (arrows), the small structure has a gap ratio $\Delta \omega_g / \omega_g = 0.6$, and the larger structure (higher velocity contrast) has $\Delta \omega_g / \omega_g = 0.9$.}
	\label{fig:2D}
\end{figure*}
We now turn to a relatively straightforward realization in two dimensions (2D) based on standard fabrication techniques and a ``membrane-in-the-middle'' optomechanical geometry \cite{Thompson2008Strong}. In Fig.~\ref{fig:2D}, we simulate the normal modes of a $100$-nm-thick \SiN membrane patterned into the hexagonal lattice inset to (a). The pad diameter $d = 16$ \um, tether width $w=1$ \um, and tether length $l=52.5$ \um; these parameters are chosen because (i) the unit cell is relatively small (meaning a millimeter-scale structure can contain many of them), (ii) the tether width is compatible with large-area photolithography (single-pad structures have recently achieved extraordinarily low dissipation rates \cite{Reinhardt2016Ultralow, Norte2016Mechanical}), (iii) the pads are large enough to not result in significant optical clipping losses \cite{Chang2012Ultrahigh, Reinhardt2016Ultralow} when positioned within a fiber cavity \cite{Jacobs2012Fiber}, and (iv) this value of $l$ maximizes the gap ratio $\Delta \omega_g/\omega_g$.

Applying Bloch-periodic boundary conditions (wave vector $\vec{k}_b$) to the parallelogram unit cell (inset) yields the dispersion relation plotted in Fig.~\ref{fig:2D}(a), following the first Brillouin zone path (``1BZ'', inset). The blue out-of-plane (OOP) modes exhibit a gap between 2.0 and 3.8 MHz.\footnote{Note we find square lattices of similar dimensions behave qualitatively similarly, but exhibit a smaller gap, due to a combination of reduced rotational symmetry (i.e. the shift between $\omega_g$ for the K and M directions is larger) and a more gradual taper between high and low wave velocities associated with the fillets (see Ref.~\cite{Reinhardt2016Ultralow} for the pad shape).} The in-plane modes (burgundy) are significantly stiffer than the OOP modes, cutting through the gap. However, since we assume the optical restoring force is applied in the OOP direction, the in-plane modes should remain orthogonal, and not play a role in OOP dynamics.

Figure \ref{fig:2D}(b) shows the normal mode frequencies for the finite crystal of (c), with a MHz-scale optical trap having Gaussian intensity profile of diameter 8 \um~applied to the central pad \cite{Chang2012Ultrahigh, Muller2015Enhanced}. Similar to the 1D model, the (i) nominally delocalized band edge mode (ii) initially localizes and then (iii) delocalizes again as it enters the upper band. This result agrees surprisingly well with the infinite 1D model (faint red curve) if we employ linear mass densities $\rho_p = 5.7$ pg/\um~and $\rho_t = 0.81$ pg/\um, estimated from the density variations of a unit cell along the ``K'' direction (a, inset), and tension $T=230~\upmu$N, estimated from the cross-sectional area of the tethers. The effective mass (magenta) again plummets to a value comparable to that of a single unit cell ($m\unit$), with $m\eff \sim  \rho_\text{SiN} t d^2 \sim 100$ pg near the middle of the gap. Other modes of the structure are trapped as well (note the subtle frequency shifts in Fig.~\ref{fig:2D}(b)), but the band edge mode is the first to benefit from a reduced $m\eff$, allowing it to quickly pull away from the band.

As shown in Fig.~\ref{fig:2D}(d), one can also achieve localization of the \emph{upper} band edge modes by applying a trap of negative strength (i.e. an ``antitrap''). Antitraps can be achieved, for example, by positioning the pad at the node of an optical standing wave where, advantageously, optical loss in the nitride is minimized, thereby alleviating the problem of excessive heating. However, for our naively-applied trap, the fundamental mode becomes unstable long before significant localization can occur, as evidenced by its immediate drop to zero frequency. On the other hand, the localized modes (inset) are torsional in nature, meaning a purely torsional trap might circumvent this problem, at the expense of increased absorption \cite{Muller2015Enhanced}.

While patterned membranes, fiber cavities, and MHz-frequency optical traps together represent a viable means of optically localizing a mechanical mode, for lower-frequency applications one may prefer to work with larger pads and free-space optics. A potential advantage of larger pads is that the increased ratio $d/w$ (assuming $w=1$ \um~remains fixed by photolithography) results in a larger velocity contrast $v_t/v_p$. This in turn creates a larger gap ratio $\Delta \omega_g/\omega_g$ and a larger difference in amplitude between neighboring pads when localized. Figure \ref{fig:2D}(e) shows the dependence of the band edge frequencies on tether length for the original pads with $d=16$ \um~(blue) and larger pads with $d=128$ \um~(red). The fractional gap $\Delta\omega_g/\omega_g$ can indeed be much higher for the large pads ($\Delta\omega_g/\omega_g = 0.9$) than for the smaller pads ($\Delta\omega_g/\omega_g=0.6$), and all mechanical frequencies of course decrease with increasing size. The trade-off for larger structures is that, in order to achieve the optimal gap, the tethers must be correspondingly lengthened to the millimeter scale (though a sub-optimal gap still exists for significantly shorter tethers). 

As a final note, we have chosen this transverse-wave geometry because it is easy to visualize and relevant to our group's experimental capabilities, but the same physics will occur in any periodic mechanical structure, provided a local optical trap can be applied.

\section{Discussion}\label{sec:discussion}

Since a typical optomechanics lab is incapable of fabricating an infinite phononic crystal, an important figure of merit is the localization length $\mathscr{L}_1$ that can be achieved with an average cavity occupancy $\bar{n}_\gamma= 1$. In the previous sections, the trap strength is normalized to avoid any dependence on a particular trapping mechanism. To get a sense of scale for a realistic system, suppose a trap is applied to a 1D crystal having the parameters discussed in Section \ref{sec:2D} by a fiber cavity \cite{Jacobs2012Fiber} of length $L=10$~\um~and finesse $\mathcal{F}=10^5$ at a wavelength $\lambda=1064$ nm, using the (stable) ``quadratic'' optomechanical coupling found near avoided crossings \cite{Lee2015Multimode, Muller2015Enhanced}. In this case, the upper limit on the per-photon spring constant $K_1$ \cite{Muller2015Enhanced} produces a normalized trap strength $\sqrt{K_1/m_\text{unit}} \sim \sqrt{16hcF/L\lambda^2 m_\text{unit}}\sim 75$~kHz\footnote{Note the ``linear'' optical spring produces a comparable trap with some cold antidamping in this system \cite{AspelmeyerEDITOR2014Cavity}.}, a localization length $\mathscr{L}_1 \sim 30$ mm ($570$ unit cells), and an effective mass $m_1 \sim 50$ ng. Remarkably, $\mathscr{L}_1$ is not a quantity naturally measured in parsecs, and this extremely low level of light should be capable of producing significant changes in the mode of a chip-scale mechanical element. To further quantify this statement, we introduce a second figure of merit, the ratio of the trapped pad's amplitude with the trap on ($y\on$) and off ($y\off$), for a fixed mechanical energy $U\tot$ stored in the mode. For a 1D crystal having $N$ unit cells in the weak-trap limit $\Delta \ll 1$, the localization length $\mathscr{L} \gg a$, and $U\tot$ can again be approximated by a geometric series, yielding
\begin{equation}\label{eq:scaling}
\frac{y\on}{y\off} \approx \sqrt{\frac{Na}{\mathscr{L}}\frac{1}{1-e^{-Na/\mathscr{L}}}}.
\end{equation}
In the ``small-crystal'' limit $Na \ll \mathscr{L}$, this expression reduces simply to $\mathscr{Y} \approx 1+\frac{1}{4}\frac{Na}{\mathscr{L}}$, highlighting the enhancement from light's combined influence over the many modes of a large crystal: for a given trap, the change scales with $N$, so larger crystals exhibit a \emph{larger} response, despite the correspondingly larger mass. This perhaps unintuitive result can be understood by noting that a larger structure has more mechanical energy to draw inward to the trapping site, or, equivalently, that the density of band-edge modes scales roughly as $N$, and the hybridization of these modes leads to a larger trapped pad amplitude. Using the same parameters as above, this effect can be quite large even for the case $\bar{n}=1$, producing a $\sim 5\%$ amplitude change in a $\sim 7$-mm-long ($N=120$) 1D crystal. On the other hand, if the localization length is \emph{smaller} than the crystal ($\mathscr{L} \ll Na$), Eq.~\ref{eq:scaling} reduces to $y\on/y\off \approx\sqrt{Na/\mathscr{L}}$. In this limit, the entire structure's mechanical energy is drawn to within a radius $\sim\mathscr{L}$ of the trap, and the resulting amplitude changes can be significantly larger (scaling even more favorably in higher dimensions). All of these effects would of course be further enhanced by using electron beam lithography to define thinner tethers.

It is currently not possible to achieve this level of \emph{in situ} control over the geometry of a solid state mechanical system, so these results provide a curious set of opportunities. For example, it is well-known that partial optical levitation improves the coherence of mechanical elements \cite{Chang2009Cavity, Romero2010Toward, Singh2010All, Chang2012Ultrahigh, Ni2012Enhancement, Muller2015Enhanced}, and the addition of spatial localization would further isolate the system from the lossy clamped boundaries \cite{Wilson2011High, Jockel2011Spectroscopy, Chakram2014Dissipation, Villanueva2014Evidence}. Along these lines, the ability to systematically \emph{tune} a mechanical mode's interaction with the boundaries (or other fabricated structures) provides access to unique studies of dissipation mechanisms, a subject of central interest to all mechanical technologies. Specifically, instead of fabricating \emph{many} (nominally) identically devices with systematically varied shapes, one could fabricate a single mechanical crystal and optically tune the mode shape to help separate the roles of bulk bending, clamping, or other structural losses. On a more fundamental side, this light-geometry interaction might aid in the pursuit of macroscopic quantum motion \cite{OConnell2010Quantum, Teufel2011Sideband, Chan2011Laser, Safavi2012Observation, Purdy2015Optomechanical, Underwood2015Measurement, Meenehan2015Pulsed}. An immediately interesting question is how a large crystal, perhaps driven to very large amplitude, might evolve under the influence of a single cavity photon, a superposition of photon states, or squeezed light. If it is possible to generate large-amplitude superpositions or other non-classical motional states with a sufficiently massive crystal, this could perhaps even provide a platform for tests of mechanisms leading to the collapse of macroscopic quantum behavior \cite{Tegmark1993Apparent}. Alternatively, this system could be used to approach the goal of quantum state transduction \cite{Tian2010Optical, Stannigel2010Optomechanical, Regal2011From, Safavi2011Proposal, Wang2012Using, Hill2012Coherent, Liu2013Electromagnetically, Andrews2014Bidirectional} from a different direction: a trap toggling the spatial extent of a mechanical mode could be used to toggle its interaction with an object at a different lattice site, e.g., a qubit or another optical resonator (perhaps operating at a very different wavelength). Finally, by intentionally adding spatial disorder (speckle) to the trapping field, it would be possible to perform controlled studies of Anderson localization \cite{Anderson1958Absence} of phonons in one or two dimensions. Ultimately, however, it is our hope that this added control will inspire a set of non-traditional optomechanics and sensing applications beyond those naively imagined here.

\section{Acknowledgments}

We thank Aashish Clerk, Yariv Yanay, Lilian Childress, Christoph Reinhardt, Simon Bernard, Maximilian Ruf, and Bogdan Piciu for helpful discussions. A.Z.B.~acknowledges support from King Abdulaziz University represented by Saudi Arabian Cultural Bureau in Canada. T.M.~acknowledges support by a Swiss National Foundation Early Postdoc Mobility Fellowship. The authors also acknowledge financial support from NSERC, FRQNT, the Alfred P.~Sloan Foundation, CFI, INTRIQ, RQMP, CMC Microsystems, and the Centre for the Physics of Materials at McGill.

\bibliography{Alles}

\end{document}